# Evolutionary search for new high-*k* dielectric materials: methodology and applications to hafnia-based oxides

Qingfeng Zeng,[a]* Artem R. Oganov,[b,c,a] Andriy O. Lyakhov,[b] Congwei Xie,[a] Xiaodong Zhang,[a,e] Jin Zhang,[a] Qiang Zhu,[b] Bingqing Wei[d,f] and Ilya Grigorenko[g]

[a]*Science and Technology on Thermostructure Composite Materials Laboratory, Northwestern Polytechnical University, Xi'an, Shaanxi 710072, PR China*
[b]*Department of Geosciences, Center for Materials by Design, and Institute for Advanced Computational Science, State University of New York, Stony Brook, NY 11794-2100, USA*
[c]*Department of Problems of Physics and Energetics, Moscow Institute of Physics and Technology, Dolgoprudny, Moscow Region 141700, Russia*
[d]*State Key Laboratory of Solidification Processing, School of Materials Science and Engineering, Northwestern Polytechnical University, Xi'an, Shaanxi 710072, PR China*
[e]*Institute of Modern Physics, Northwest University, Xi'an, Shaanxi 710069, PR China*
[f]*Department of Mechanical Engineering, University of Delaware, Newark, DE 19716, USA*
[g]*Physics Department, New York City College of Technology, The City University of New York, Brooklyn, NY 11201, USA*

*E-mail: qfzeng@nwpu.edu.cn*

## Abstract

High-*k* dielectric materials are important as gate oxides in microelectronics and as potential dielectrics for capacitors. In order to enable computational discovery of novel high-*k* dielectric materials, we propose a fitness model (energy storage density) that includes the dielectric constant, bandgap, and intrinsic breakdown field. This model, used as fitness function in conjunction with first-principles calculations and global optimization evolutionary algorithm USPEX, efficiently leads to practically important results. We found a number of high-fitness structures of $SiO_2$ and $HfO_2$, some of which correspond to known phases and some are new. The results allow us to propose characteristics (genes) common to high-fitness structures – these are the coordination polyhedra and their degree of distortion. Our variable-composition searches in the $HfO_2$-$SiO_2$ system uncovered several high-fitness states. This hybrid algorithm opens up a new avenue of discovering novel high-*k* dielectrics with both fixed and variable compositions, and will speed up the process of materials discovery.

## 1. Introduction

$SiO_2$ as a gate oxide has emerged as one of the key bottlenecks in downscaling a wide variety of devices considering its unacceptably high leakage current level originating from electron tunneling(Iwai & Ohmi, 2002, Sasaki *et al.*, 1996). Thus, finding superior alternatives, which have better electric characteristics and capable of forming a stable interface with the substrate, has become a major challenge. Among the various candidates for replacing $SiO_2$ as a gate oxide are $ZrO_2$, $HfO_2$, $TiO_2$, $Al_2O_3$, $Ta_2O_5$, and $Si_3N_4$(Campbell *et al.*, 1997, Gerritsen *et al.*, 2005, Kadoshima *et al.*, 2003, Nahar *et al.*, 2007), but it is hafnia（$HfO_2$）that has attracted the greatest interest for future sub-0.1$\mu$m gate-dielectric thickness in complementary metal-oxide semiconductor (CMOS)(Wilk *et al.*, 2001, Kingon *et al.*, 2000, Robertson, 2004) because of its excellent static dielectric constant ($k$=22~25), bandgap ($E_g$=5.5~6.0 eV), and the electric breakdown field ($E_{bd}$=0.39~0.67 V/nm)(Nahar *et al.*, 2007). At the same time, hafnium-based oxides can be made structurally stable on silicon during fabrication and operation by sufficiently increasing both the crystallization temperature and resistance to oxygen diffusion, without significantly compromising the high dielectric permittivity of $HfO_2$(Choi *et al.*, 2011).

It seems reasonable to merely search for materials with the highest possible dielectric constant materials. However, since the barrier height tends to decrease with increasing dielectric constant(Wilk *et al.*, 2001), the direct tunneling leakage is not solely or exclusively determined by the value of *k*(Wu *et al.*, 2006). Consequently, a carefully designed balance needs to be achieved for the static dielectric constant, bandgap, and the electric breakdown field(Kawamoto *et al.*, 2001), given that it is impossible to simultaneously maximize $k$, $E_g$ and $E_{bd}$, at least in simple binary oxides(Choi *et al.*, 2011). Crystal structure determines stability of materials and properties such as the dielectric constant and bandgap(Robertson, 2006, Caravaca & Casali, 2005), and therefore we focus our search on crystal structure prediction.

With this motivation, we perform hybrid optimization(Lyakhov & Oganov, 2011, Zhu *et al.*, 2011)– i.e. we search for the globally optimal (with respect to fitness, i.e. the desired physical property described below) atomic configuration, considering only structures corresponding to local minima of the energy, i.e. structures that can exist as stable or metastable. Combination of local energy optimization and global fitness optimization makes us call it "hybrid".

In this study, we adopt a fitness carefully balancing the pertinent variables, namely $k$, $E_g$, and $E_{bd}$, ensuring the selected materials are highly scalable and usable for the next generation of microelectronic devices. All the relevant properties are calculated using the first-principles calculations. This approach has been implemented in the USPEX (Universal Structure Prediction: Evolutionary Xtallography) code(Glass *et al.*, 2006). Present paper is organized as follows. In Section 2, we describe the fitness model. In Section 3, we describe the calculation methods and parameters. In Section 4, we discuss the application of this approach to hafnia and hafnia-silica systems. Finally, in Section 5 we summarize the results and draw conclusions.

## 2. Fitness Model

Under external electric field, electronic clouds and ions slightly shift from their equilibrium positions, creating dielectric polarization that results in the storage and dissipation of electromagnetic energy in materials. Hence, dielectric properties are recognized to be of central importance for explaining various phenomena in electronics, optics, and solid state physics(Aspnes & Studna, 1983, Zhao & Vanderbilt, 2002). A dielectric is also called an insulator because it causes electrical obstruction, due to which a gate oxide enables the gate to acquire a high transverse electric field that can strongly modulate the conductance of the channel in an electronic device(Hybertsen & Louie, 1987). Materials with high dielectric constants are also used in capacitors. The measure of the electromagnetic energy stored within a dielectric is given by

$$E = \frac{1}{2} C U^2 \qquad (1)$$

where $C$ is capacitance (in F) and $U$ is the electric field (in V). This equation shows that the higher the capacitance and the electric field, the greater energy storage, and the better performance of the devices.

A gate dielectric film forms a parallel plate capacitor with capacitance $C = \varepsilon_0 kA/d$, where $A$ is the area of the film, $d$ thickness, $\varepsilon_0$ the absolute permittivity of the vacuum (a constant equal to $8.854\times10^{-12}$ F/m), and $k$ the relative permittivity. Relative permittivity, more usually called 'dielectric constant', is a dimensionless quantity (a ratio of the absolute permittivity of a dielectric medium to that of the vacuum). All materials will therefore have dielectric constants greater than 1. In general, promising high-*k* materials should have dielectric constants between 10 and 30(Choi *et al.*, 2011), about one order of magnitude higher than that of alpha-quartz. One way to increase the capacitance is to decrease $d$, which will eventually result in quantum-mechanical tunneling when a film scales down to 1 nm. Increasing the area of the film will not allow miniaturization of the devices, therefore, for a dielectric film of limited area and thickness to effectively suppress the leakage current, we need to find materials with high dielectric constants.

Another parameter related to the energy storage is the electric field that can be applied to the capacitor without causing dielectric breakdown. The maximum breakdown field is related to the

bandgap(Wang, 2006). Experimental tests provide a universal relationship(Wang, 2006, Ieong *et al.*, 2004) between intrinsic breakdown field $E_{BI}$ and the bandgap $E_g$: $E_{BI} = 1.36 \times (E_g / E_{gc})^{\alpha}$ (V/nm), where $E_{gc} = 4.0$ eV, which is the critical bandgap value separating materials into semiconductors (below $E_{gc}$) and insulators (above $E_{gc}$), $\alpha = 3$ for semiconductors, and $\alpha = 1$ for insulators. It is desirable to have high-*k* materials with bandgaps greater than 4.0 eV, because in that case the breakdown field can be very high. This conclusion meets the reported criteria for high-*k* gate materials(Ieong *et al.*, 2004).

Consequently, the final fitness model should be given as maximizing the energy density

$$\max F_{ED} = \frac{1}{2}\varepsilon_0 k E_{BI}^2 = 8.1882\ \text{J/cm}^3 \times k\left(E_g / E_{gc}\right)^{2\alpha} \quad (2)$$

With this new fitness descriptor, we can simultaneously account for the dielectric constant, bangap, and breakdown field during optimization, in a rational and comprehensive way. Remarkably, the same fitness can be used to search for optimal dielectric materials for capacitors and gate oxide materials. The dielectric constant is a tensorial quantity closely related to crystal structure, electronic structure and lattice dynamics – but since most dielectrics are used in the polycrystalline form, here we use the orientationally averaged static dielectric constant.

## 3. Structure prediction and search for new dielectric materials

Computational materials discovery includes as its integral part crystal structure prediction(Oganov & Glass, 2006, Oganov & Glass, 2008). In our calculations, the dielectric constant and bandgap are evaluated for fully relaxed structures. These parameters are used to get the value of the fitness descriptor for global optimization. Here we use the concept of hybrid optimization – where global optimization with respect to a given physical property (fitness) is done over the set of local minima of the total energy (or another relevant thermodynamic potential)(Lyakhov & Oganov, 2011, Zhu *et al.*, 2011). Global optimization was done using the USPEX code(Oganov & Glass, 2006, Oganov *et al.*, 2011, Lyakhov *et al.*, 2013).

Total energy calculations and structure relaxations were done using the generalized gradient approximation(Perdew *et al.*, 1996) as implemented in the VASP code(Kresse & Furthmüller, 1996). Our *ab initio* calculations used the all-electron projector augmented wave (PAW method) (Blöchl, 1994) ([He] core with 1.52 a.u. radius for O, [Xe] core with radius 3.0 a.u. for Hf, [Ne] core with 1.9 a.u. radius for Si), plane-wave basis sets with the kinetic energy cutoff of 600 eV, and Monkhorst-Pack meshes for Brillouin zone sampling with reciprocal space resolution of $2\pi \times 0.06$ Å$^{-1}$. These settings are sufficient for excellent convergence of the total energy, stress tensor, dielectric constant, and bandgap. DFT bandgaps are known to be underestimated, and more accurate values can be obtained (though at a much greater computational cost) by other methods, e.g. the GW approximation(Shishkin & Kresse, 2006, 2007) or hybrid functionals(Da Silva *et al.*, 2007). Here we use DFT bandgaps as approximate and useful guide values.

For optimal high-*k* gate materials, in addition to optimal physical properties, we also need to ensure a good interface compatibility with the silicon substrate(Lee *et al.*, 1999, Lee *et al.*, 2000). Consequently, the structures of mixed hafnium-silicon oxides ($Hf_{1-x}Si_xO_2$, which can be considered as $(HfO_2)_{1-x}(SiO_2)_x$ pseudobinary compounds) are worth of investigating through this new fitness model.

The variable-composition optimization(Oganov *et al.*, 2011) is a particularly powerful feature of USPEX, through which we can get the crystal structure information not only of end members $HfO_2$ and $SiO_2$ but also of pseudobinary $HfO_2$-$SiO_2$ compounds and mixtures, allowing rapid screening of the compositional space through artificial intelligence. In order to find out the character of the crystal structure and its relationship with dielectric property, we explore $HfO_2$, $SiO_2$ and $HfO_2$-$SiO_2$. In variable-composition searches, any $HfO_2$-$SiO_2$ compositions are allowed, under constraint that the total number of atoms be in the range 3-30 atoms/cell. The first generation contains 50 candidate structures, all subsequent generations contain 40 structures. 50% of the new generation is produced by heredity, 20% by softmutation, 20% by permutation, and 10% by transmutation.

Calculations were allowed to run for at most 30 generations, and were stopped if for 10 generations the best found solution did not change. The structures were visualized by VESTA(Momma & Izumi, 2011).

## 4. Results and discussion

Hafnia has three main polymorphs: monoclinic (*P2₁/c*), tetragonal (*P4₂/nmc*) and cubic (*Fm-3m*)(Quintard *et al.*, 2002, Lee *et al.*, 2008). Monoclinic phase has baddeleyite-type structure and is thermodynamically stable at normal conditions, while the tetragonal polymorph (distorted fluorite-type structure, the ideal fluorite structure corresponding to the cubic polymorph) has the largest dielectric permittivity among the three phases. In our crystal structure searches, all these three well-known phases were found. The tetragonal phase has the highest fitness value (604.04 J/cm$^3$) since it has the largest dielectric constant (57.52) and the second largest bandgap (4.53 eV) among the three phases. Besides the three well-known phases, we have found several other interesting metastable phases with high fitness values. The most interesting crystal structures of the $HfO_2$ polymorphs discovered by USPEX are given in Fig. 1 and Table 1.

The $P2_1/m$ (Fig.1b) phase is a newly predicted structure with a slightly higher enthalpy (0.0277 eV/atom) compared to the $P2_1/c$ phase and a significantly improved fitness (174.24 J/cm$^3$ instead of 121.59 J/cm$^3$, 43% enhancement in energy storage). Table 2 shows that the dielectric constant and the corresponding fitness of the tetragonal-$HfO_2$ ($P4_2/nmc$) are much higher than those of other polymorphs. The tetragonal phase is the distortion of the fluorite structure (the cubic phase) and have 8-fold coordination of Hf atoms; the $P2_1/c$ structure is topologically quite different, it belongs to the baddeleyite structure type and has 7-fold coordination of Hf; the *P-1* and $P2_1/m$ structures have 6-fold coordination of Hf. Analyzing our results, we see that the coordination number of Hf is an appropriate predictor for high-*k* materials – and thus can be considered a “gene” determining high fitness. Generally, we find that the dielectric constant increases with the coordination number, which is easy to understand as increasing the coordination number leads to decreasing bond strength and corresponding vibrational frequencies, which makes ionic polarization easier. This is also the reason why the dielectric constants are usually much higher for $HfO_2$ than for $SiO_2$.

Some phases (e.g. *Pnma* and $Pca2_1$) have similar dielectric constants (21.42 and 20.12), while they have drastically different fitness values (48.47 J/cm$^3$ and 191.28 J/cm$^3$) because of their difference in terms of bandgaps (3.23 eV and 4.31 eV). Thus, we need to trade-off the $k$ and $E_g$ to get a higher $F_{ED}$. Phases with bandgaps above 4.0 eV have intrinsic breakdown voltages exceed 1.36 V/nm, i.e. they meet the working voltage requirement in Intel’s first 22nm-based 3-D transistors and other CMOS devices(Batude *et al.*, 2009). We note, however, that bandgaps (and breakdown voltages) based on density-functional calculations are always underestimated, usually by 10-40%. Thus, our predicted fitnesses are underestimated, and the actual fitness values will be even better, thus making our predictions safe for practical applications.

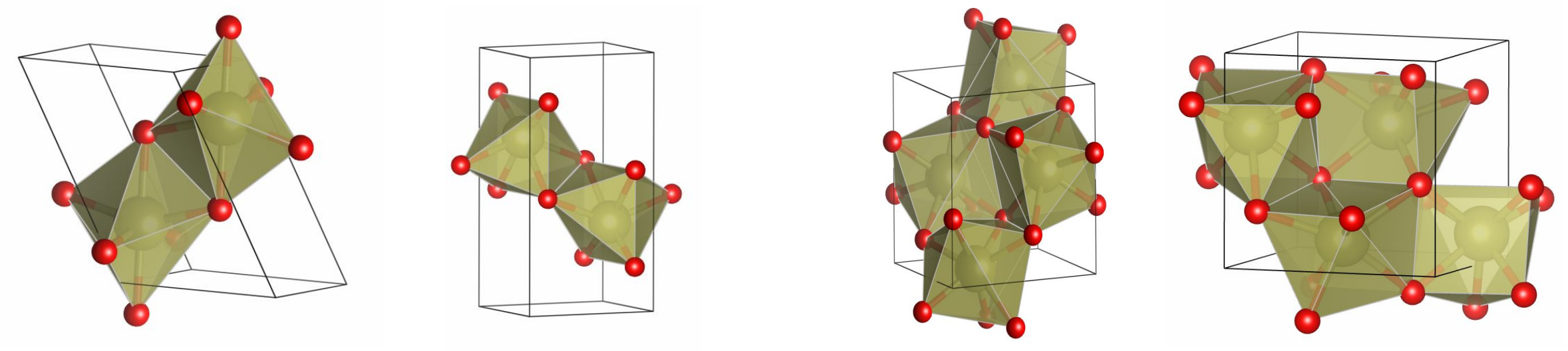

(a) *P-1* (b) $P2_1/m$ (c) $P2_1/c$ (d) $Pca2_1$

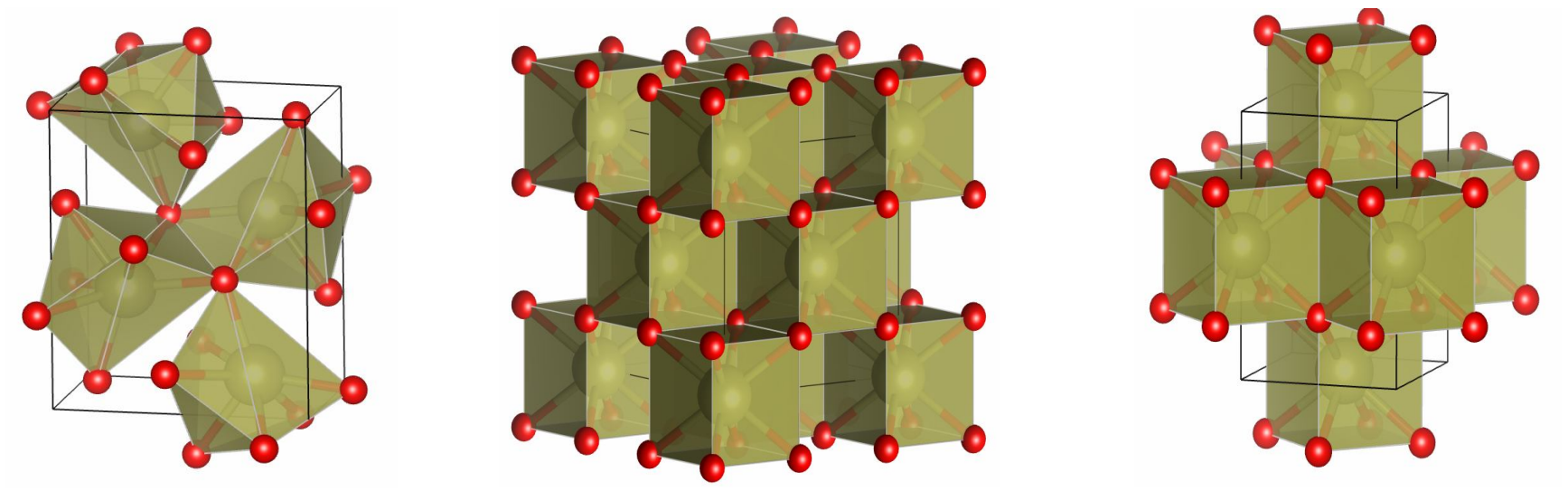

(e) *Pnma* (f) *Fm-3m* (g) *P4$_2$/nmc*

Fig. 1 The main crystal structures of $HfO_2$ discovered by USPEX

Table 1 Lattice parameters of the $HfO_2$ polymorphs discovered by USPEX

| Compound | Relative enthalpy (eV/atom) | Space group (No.) | Lattice constants (Å) Present | Reported | Atom (Wyckoff position), coordinates atom | x | y | z |
|---|---|---|---|---|---|---|---|---|
| triclinic-$HfO_2$ | 0.0428 | *P-1*<br>(2) | *a*=3.894<br>*b*=5.848<br>*c*=3.895<br>α=70.614<br>β=89.993<br>γ=109.400 | | Hf (2i)<br>O1 (2i)<br>O2 (2i) | 0.8746<br>0.0769<br>0.3261 | 0.7500<br>0.1526<br>0.6526 | 0.3746<br>0.1740<br>0.4228 |
| monoclinic-$HfO_2$ | 0.0277 | *P2$_1$/m*<br>(11) | *a*=5.376<br>*b*=3.427<br>*c*=3.823<br>β=104.474 | | Hf (2e)<br>O1 (2e)<br>O2 (2e) | 0.7701<br>0.4185<br>0.9357 | 0.2500<br>0.2500<br>0.2500 | 0.8385<br>0.2352<br>0.2660 |
| monoclinic-$HfO_2$ | 0 | *P2$_1$/c*<br>(14) | *a*=5.085<br>*b*=5.159<br>*c*=5.262<br>β=99.702 | *a*=5.118[a]<br>*b*=5.185<br>*c*=5.284<br>β=99.352 | Hf (4e)<br>O1 (4e)<br>O2 (4e) | 0.2766<br>0.0678<br>0.4513 | 0.0468<br>0.3320<br>0.7417 | 0.2091<br>0.3462<br>0.4757 |
| orthorhombic-$HfO_2$ | 0.0209 | *Pca2$_1$*<br>(29) | *a*=5.107<br>*b*=4.900<br>*c*=4.920 | | Hf (4a)<br>O1 (4a)<br>O2 (4a) | 0.9670<br>0.1303<br>0.7338 | 0.7317<br>0.0734<br>0.4593 | 0.1280<br>0.2632<br>0.8735 |
| orthorhombic-$HfO_2$ | 0.0904 | *Pnma*<br>(62) | *a*=5.386<br>*b*=3.213<br>*c*=6.282 | | Hf (4c)<br>O1 (4c)<br>O2 (4c) | -0.2453<br>0.0241<br>-0.3599 | 0.2500<br>0.2500<br>0.2500 | -0.1122<br>0.3396<br>-0.4254 |
| cubic-$HfO_2$ | 0.0694 | *Fm-3m*<br>(225) | *a*=5.029 | *a*=5.115[b] | Hf (4a)<br>O (8c) | 0<br>0.7500 | 0<br>-0.7500 | 0<br>-0.7500 |
| tetragonal-$HfO_2$ | 0.0457 | *P4$_2$/nmc*<br>(137) | *a*=3.557<br>*c*=5.146 | *a*=3.58[c]<br>*c*=5.20 | Hf (2b)<br>O (4d) | 0.5000<br>0.2500 | 0.5000<br>0.2500 | 0<br>0.2100 |

[a] (Kresse & Furthmüller, 1996)
[b] (Quintard *et al.*, 2002)
[c] (Blöchl, 1994)

Table 2 Average static dielectric constant $k$, bandgap $E_g$, intrinsic breakdown field $E_{BI}$ and $F_{ED}$ of the $HfO_2$ polymorphs

| Compound | Space group (Coord. no.) | Dielectric constant $k$ Present | Reported | Bandgap $E_g$ (eV) Present | Reported | $E_{BI}$ (V/nm) | $F_{ED}$ (J/cm$^3$) |
|---|---|---|---|---|---|---|---|
| triclinic-$HfO_2$ | *P-1*<br>(6) | 12.76 | | 4.29 | | 1.46 | 120.28 |
| monoclinic-$HfO_2$ | *P2$_1$/m*<br>(6) | 14.65 | | 4.82 | | 1.64 | 174.24 |
| monoclinic-$HfO_2$ | *P2$_1$/c*<br>(7) | 16.76 | 16[a]<br>15-17[b]* | 3.92 | 3.95[c]<br>5.70[d]* | 1.28 | 121.59 |
| orthorhombic-$HfO_2$ | *Pca2$_1$*<br>(7) | 20.12 | | 4.31 | | 1.47 | 191.28 |
| orthorhombic-$HfO_2$ | *Pnma*<br>(7) | 21.42 | | 3.23 | | 0.71 | 48.47 |
| cubic-$HfO_2$ | *Fm-3m*<br>(8) | 30.46 | 29[a] | 3.86 | 3.55[c] | 1.22 | 201.68 |
| tetragonal-$HfO_2$ | *P4$_2$/nmc*<br>(8) | 57.52 | 70[a] | 4.53 | 4.36[c] | 1.54 | 604.04 |

* Experimental data
[a] (Zhao & Vanderbilt, 2002)
[b] (Kukli *et al.*, 1996)
[c] (Jiang *et al.*, 2010)
[d] (Bersch *et al.*, 2008)

Similar trends are found for $SiO_2$; the most interesting structures discovered by USPEX are shown in Fig. 2. Again, we find that the coordination number is a suitable (though only in the first crude approximation) predictor (gene) of high-*k* materials. Here, all structures can be grouped into two classes – (1) with tetrahedral coordination of Si and (2) with octahedral coordination of Si. Just as for $HfO_2$, here we see that structures with higher coordination have higher dielectric constants: compare alpha-quartz ($k$=3.9) and stishovite ($k$=10.55). The band gap slightly decreases in the same direction (from 5.57 eV to 5.22 eV, due to the increased occupancy of d-orbitals of Si), but still fitness is greatly improved for the octahedral structures. Stishovite has the highest fitness (146.98 J/cm$^3$, compared with 75.90 J/cm$^3$ for alpha-quartz, almost double improvement of energy storage) and meets the lowest boundary for high-*k* materials ($k$=10~30). Although the synthesis of the structures with octahedral Si usually require high pressure condition, stishovite can be quenched to ambient condition once being synthesized. So far, only one silicate with such coordination is formed at atmospheric pressure, thaumasite $Ca_3Si(OH)_6(CO_3)(SO_4)\cdot 12H_2O$. Note also that fitness values for silica polymorphs are generally lower than those for phases of hafnia; hence, alloying with hafnia can be expected to improve performance of silica-based dielectrics. In our following investigation of $HfO_2$-$SiO_2$, we will see the characteristic genes from both $HfO_2$ and $SiO_2$ which contribute to increase their fitness.

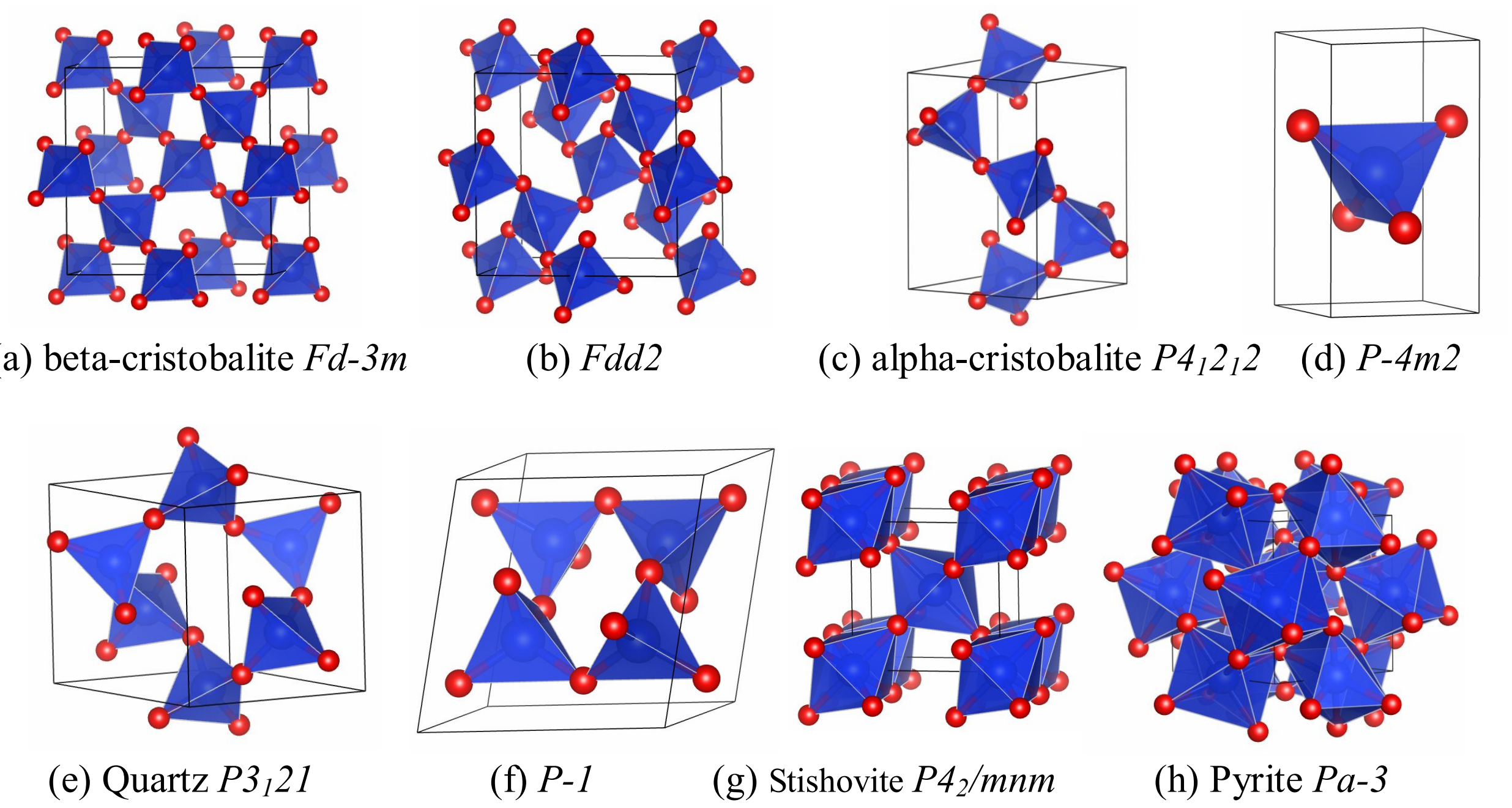

Fig. 2 The main crystal structures of $SiO_2$ discovered by USPEX

Table 3 Lattice parameters of the $SiO_2$ polymorphs discovered by USPEX

| Compound | Relative enthalpy (eV/atom) | Space group (No.) | Lattice constants (Å) Present | Lattice constants (Å) Reported | Atom (Wyckoff position), coordinates: Atom | x | y | z |
|---|---|---|---|---|---|---|---|---|
| tetragonal- $SiO_2$ | 0.2231 | *P-4m2* (115) | $a$=2.8120 $c$=4.8540 | | Si(1c) | 0.5000 | 0.5000 | 0.5000 |
| | | | | | O(2g) | 0 | 0.5000 | 0.3203 |
| triclinic-$SiO_2$ | 0.0678 | *P-1*<br>(2) | *a*=4.9940<br>*b*=5.4810<br>*c*=5.5110<br>α=102.9920<br>β=92.8180<br>γ=90.1320 | | Si1(2i)<br>Si2(2i)<br>O1(2i)<br>O2(2i)<br>O3(2i)<br>O4(2i) | -0.3016<br>0.1803<br>0.1256<br>-0.2280<br>0.3824<br>0.2392 | 0.2889<br>0.2105<br>-0.2358<br>0.0767<br>0.2630<br>0.4237 | -0.3301<br>0.3287<br>-0.4202<br>-0.1694<br>-0.4206<br>0.1702 |
| tetragonal-$SiO_2$<br>(alpha-cristobalite) | 0.0029 | *P4₁2₁2*<br>(92) | *a*=5.0233<br>*c*=7.0008 | a=4.9028[a]<br>c=6.7782 | Si(4a)<br>O(8b) | 0.3007<br>0.2384 | 0.3007<br>0.1078 | 0<br>0.1813 |
| trigonal-$SiO_2$<br>(alpha-quartz) | 0 | *P3₁21*<br>(152) | *a*=4.9508<br>*c*=5.4494<br>γ=120 | a=4.9140[b]<br>c=5.4060<br>γ=120 | Si(3a)<br>O(6c) | -0.4683<br>-0.1408 | 0<br>0.2707 | 1/3<br>0.4490 |
| orthorhombic-$SiO_2$ | 0.0064 | *Fdd2*<br>(43) | *a*=6.8180<br>*b*=6.7660<br>*c*=7.3880 | | Si(8a)<br>O(16d) | 0.2500<br>-0.0625 | 0.2500<br>-0.1869 | -0.3271<br>0.5476 |
| cubic-$SiO_2$<br>(beta-cristobalite) | 0.0187 | *Fd-3m*<br>(227) | *a*=7.4640 | a=7.132[c] | Si(8b)<br>O(16d) | 0.2500<br>0.1250 | -0.2500<br>-0.1250 | 0.2500<br>0.3750 |
| tetragonal-$SiO_2$<br>(stishovite) | 0.0576 | *P4₂/mnm*<br>(136) | *a*=4.1910<br>*c*=2.6850 | a=4.0440[c]<br>c=2.6190 | Si(2a)<br>O(4f) | 0<br>-0.3057 | 0<br>-0.3057 | 0<br>0 |
| cubic-$SiO_2$<br>(pyrite-type) | 0.4081 | *Pa-3*<br>(205) | *a*=4.4530 | | Si(4b)<br>O(8c) | 0.5000<br>-0.1558 | 0.5000<br>-0.1558 | 0.5000<br>-0.1558 |

[a] (Downs & Palmer, 1994)
[b] (Proffen *et al.*, 2005)
[c] (Yamanaka, 2005)

Table 4 Average static dielectric constant $k$, band gap $E_g$, intrinsic breakdown field $E_{BI}$ and $F_{ED}$ of the $SiO_2$ polymorphs

| Compound | Space group (Coord. no.) | Dielectric constant $k$ | | Bandgap $E_g$ (eV) | | $E_{BI}$ (V/nm) | $F_{ED}$ (J/cm$^3$) |
|---|---|---|---|---|---|---|---|
| | | Present | Reported | Present | Reported | | |
| cubic-$SiO_2$ (beta-cristobalite) | *Fd-3m* (4) | 3.66 | | 5.38 | 6.79[c] | 1.83 | 54.12 |
| orthorhombic-$SiO_2$ | *Fdd2* (4) | 4.12 | | 5.87 | | 2.0 | 72.79 |
| tetragonal-$SiO_2$ (alpha-cristobalite) | *P4₁2₁2* (4) | 4.14 | 4.30[a] | 5.44 | 5.68[c] | 1.85 | 62.72 |
| tetragonal- $SiO_2$ | *P-4m2* (4) | 4.64 | | 5.07 | | 1.73 | 61.08 |
| trigonal-$SiO_2$ (alpha-quartz) | *P3₁21* (4) | 4.77 | 4.83[a]<br>3.9[b]* | 5.57 | 5.59[c]<br>8.9[d]* | 1.90 | 75.90 |
| triclinic-$SiO_2$ | *P-1* (4) | 4.97 | | 5.41 | | 1.84 | 74.59 |
| tetragonal-$SiO_2$ (stishovite) | *P4₂/mnm* (6) | 10.55 | 10.33[a] | 5.22 | 5.15[c] | 1.77 | 146.98 |
| cubic-$SiO_2$ (pyrite-type) | *Pa-3* (6) | 14.42 | | 4.30 | | 1.46 | 136.66 |

* Experimental data
[a] (Rignanese *et al.*, 2002)
[b] (McPherson *et al.*, 2003)
[c] (Xu & Ching, 1991)
[d] (Bersch *et al.*, 2008)

Pseudo-alloys $HfO_2$-$SiO_2$ show the same characteristic genes of high-$k$ behavior as $HfO_2$ and $SiO_2$. Different coordinations of Hf and Si coexist in these compounds (Fig. 3). For the pseudobinary compounds, the dielectric constants are somewhere in between the values for pure $HfO_2$ and $SiO_2$ (Table 6). Only one compound, $HfSiO_4$ (space group *I4₁/amd*), is found to be thermodynamically stable – and it indeed corresponds to the known phase that even exists in nature as mineral hafnon. Some metastable phases are also worth noting. For example, *P-43m* and *I-42m* forms of $Hf_3Si_1O_8$ – where the tetragonal *I-42m* form is obtained from *P-43m* by Hf-Si swap (permutation), which improves the enthalpy of formation from oxides from 0.0352 eV/atom to 0.0304 eV/atom, the orientation-averaged dielectric constant from 18.01 to 21.43, the bandgap from 4.42 eV to 4.64 eV – as a result, its fitness improves by 32%. This illustrates how effectively the variation operators of USPEX (such as permutation, or atomic swap) can improve the fitness.

The compositional dependences of enthalpy of formation and energy density are illustrated in

Fig. 4. The highest $F_{ED}$ is picked at each composition for $HfO_2$-$SiO_2$ system (including crystalline and random structures). Tetragonal $HfO_2$ (*P4₂/nmc*) has the highest fitness (604.04 J/cm$^3$) because of its much higher permittivity (57.52) and satisfactory bandgap (4.53 eV). $Hf_3SiO_8$ (*I-42m*) has the highest fitness (236.56 J/cm$^3$) among the pseudobinary $HfO_2$-$SiO_2$ compounds, because of its highest dielectric constant (21.43). Hafnon ($HfSiO_4$, space group $I4_1/amd$) has nearly the same fitness (235.42 J/cm$^3$) because of its widest bandgap (5.44 eV). Stishovite ($SiO_2$, space group $P4_2/mnm$) has the highest fitness (146.98 J/cm$^3$) among silica polymorphs because of its relatively high permittivity (10.55) and bandgap (5.22 eV). The relation between composition and energy density appeals to be intriguing (recall that the two physical quantities determining it, $E_g$ and $k$, display quite different variation with composition). High concentration of $HfO_2$ does not necessarily result in high energy storage. As an example of a disordered structure, we take $Hf_{0.9}Si_{0.1}O_2$ ($Hf_9SiO_{20}$) with a relatively large unit cell containing 30 atoms; its dielectric permittivity is relatively high (22.11), but its low $E_g$ (3.02 eV) results in very low fitness (33.93 J/cm$^3$). Ordered phases seem to be superior in terms of their fitness. Among the pseudobinary compounds, the best fitness values are possessed by $Hf_{0.5}Si_{0.5}O_2$ ($I4_1/amd$) and $Hf_{0.75}Si_{0.25}O_2$ (*I-42m*); their fitness is ~3 times higher than that of $SiO_2$ alpha-quartz. Clearly, further improvements are possible by considering other alloying systems. The methodology and principles presented here allow systematic search for such improved materials.

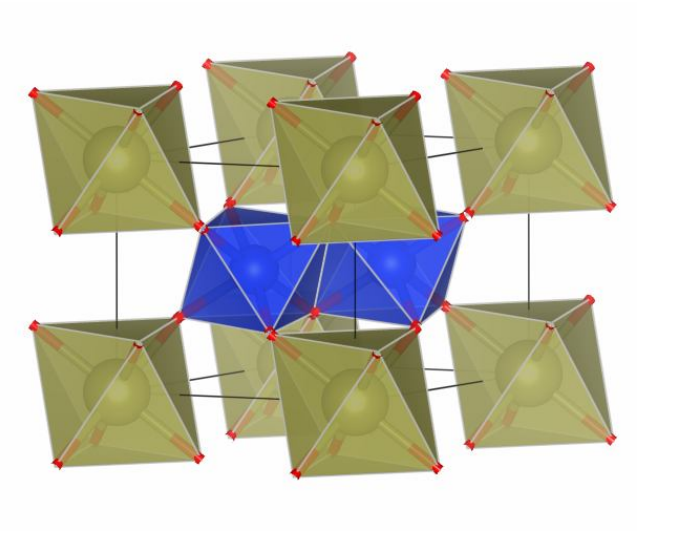

(a) *P-31m* ($Hf_{0.33}Si_{0.67}O_2$)

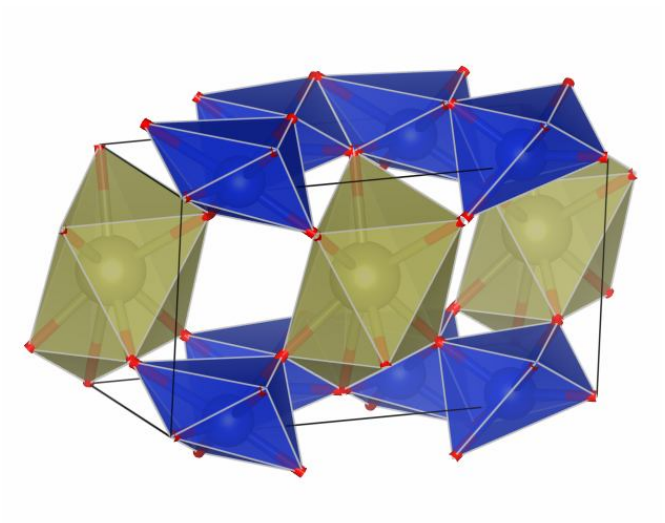

(b) *Cm* ($Hf_{0.33}Si_{0.67}O_2$)

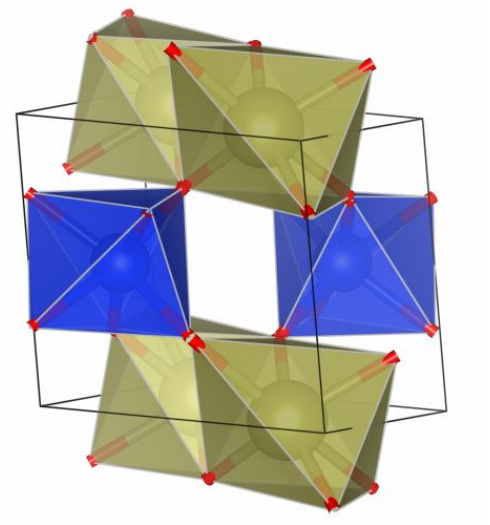

(c) *P2/c* ($Hf_{0.5}Si_{0.5}O_2$)

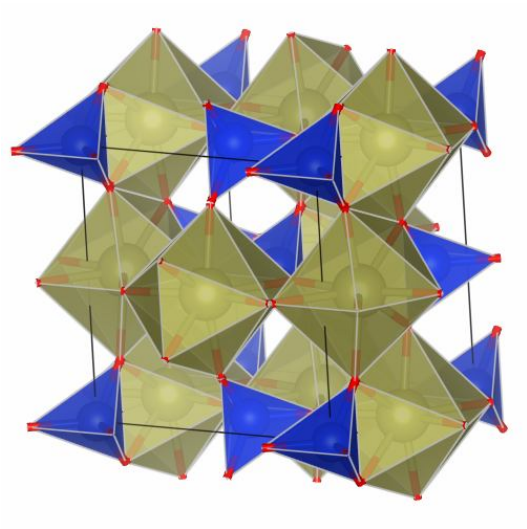

(d) $I4_1/amd$ ($Hf_{0.5}Si_{0.5}O_2$)

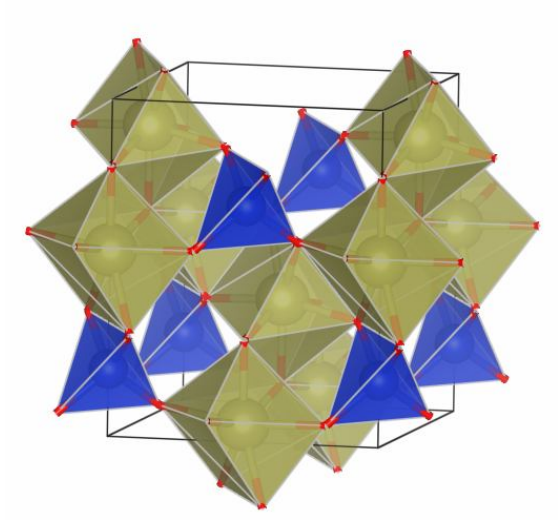

(e) *C2* ($Hf_{0.67}Si_{0.33}O_2$)

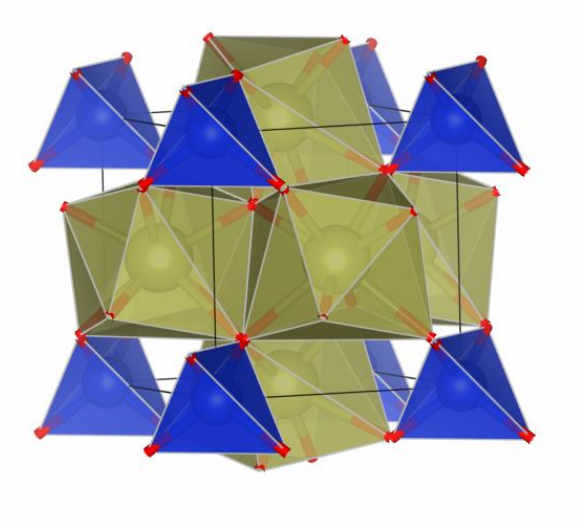

(f) *P-43m* ($Hf_{0.75}Si_{0.25}O_2$)

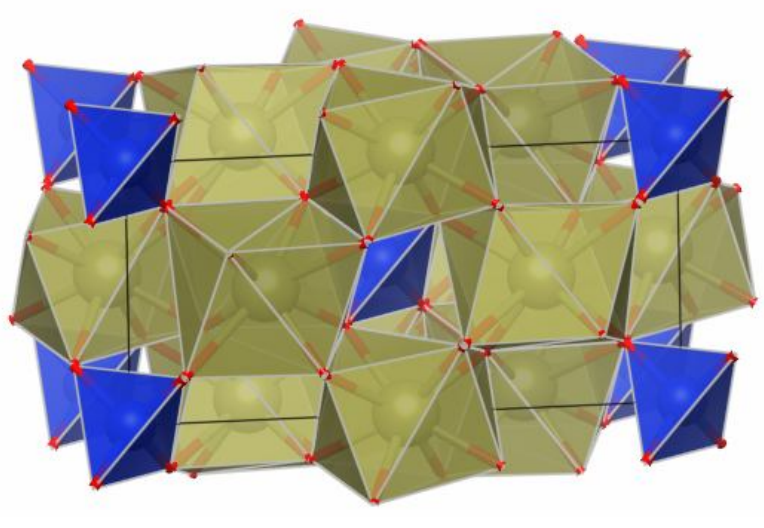

(g) *I-42m* ($Hf_{0.75}Si_{0.25}O_2$)

Fig. 3 The main crystal structures of $HfO_2$-$SiO_2$ compounds discovered by USPEX

Table 5 Lattice parameters of the $HfO_2$-$SiO_2$ polymorphs discovered by USPEX

| Compound | Enthalpy of formation (eV/atom) | Space group (No.) | Lattice constants (Å) | | Atom (Wyckoff position), coordinates | | | |
|---|---|---|---|---|---|---|---|---|
| | | | Present | Reported | Atom | x | y | z |
| $Hf_1Si_2O_6$ | 0.0417 | *P-31m* (162) | *a*=4.6300 | | Hf (1a) | 0 | 0 | 0 |
| | | | *c*=4.4300 | | Si (2d) | 2/3 | 1/3 | 1/2 |
| | | | γ=120 | | O (6k) | 0.3565 | 0.3565 | 0.7220 |
| $Hf_2Si_4O_{12}$ | 0.0771 | *Cm* (8) | | | Hf (2a) | -0.0211 | 0 | 0.0257 |
| | | | *a*=4.6810 | | Si (4b) | -0.9179 | 0.1685 | -0.4926 |
| | | | *b*=7.9790 | | O1(2a) | -0.7396 | 0 | -0.2641 |
| | | | *c*=5.8790 | | O2(2a) | -0.0350 | 0 | 0.2895 |
| | | | β=128.994 | | O3(4b) | 0.5049 | 0.1751 | 0.2923 |
| | | | | | O4(4b) | -0.3431 | 0.1826 | -0.2775 |
| $Hf_2Si_2O_8$ | 0.0923 | *P2/c* (13) | *a*=4.5904 | | Hf (2e) | 0 | 0.2789 | 0.2500 |
| | | | *b*=5.0226 | | Si(2f) | 0.500 | -0.1712 | 0.2500 |
| | | | *c*=4.8107 | | O1 (4g) | 0.2748 | -0.0787 | -0.0186 |
| | | | β=81.5289 | | O2(4g) | 0.2460 | -0.4015 | 0.3957 |
| $Hf_4Si_4O_{16}$ | -0.0378 | *I4₁/amd* (141) | *a*=6.5674 | *a*=6.61[a] | Hf (4b) | 0.5 | 1.0 | 0.25 |
| | | | *c*=5.9535 | *c*=5.97 | Si (4a) | 0 | 0.5 | 0.25 |
| | | | | | O (16h) | 0.31519 | 0.5 | 0.6831 |
| $Hf_4Si_2O_{12}$ | 0.0286 | *C2* (5) | | | Hf1 (2a) | 0 | 0.4473 | 0 |
| | | | *a*=6.4298 | | Hf2 (2b) | 0 | 0.1309 | 0.5 |
| | | | *b*=8.2462 | | Si (2a) | 0 | -0.1993 | 0 |
| | | | *c*=3.8816 | | O1 (4c) | 0.3539 | 0.4208 | -0.1664 |
| | | | β=68.8113 | | O2 (4c) | 0.3429 | 0.1627 | 0.2865 |
| | | | | | O3 (2a) | 0 | 0.2001 | 0 |
| | | | | | O4 (2b) | 0 | 0.3836 | 0.5000 |
| $Hf_3SiO_8$ | 0.0352 | *P-43m* (215) | | | Hf (3c) | 0.5000 | 0 | 0.5000 |
| | | | *a*=4.8900 | | Si (1a) | 0 | 0 | 0 |
| | | | | | O1 (4e) | 0.1992 | 0.8008 | 0.1992 |
| | | | | | O2 (4e) | 0.3024 | 0.6976 | 0.6976 |
| $Hf_6Si_2O_{16}$ | 0.0304 | *I-42m* (121) | | | Hf1 (2a) | 0 | 0 | 0 |
| | | | *a*=4.8322 | | Hf2 (4d) | 0 | 0.5000 | 0.2500 |
| | | | *c*=10.1647 | | Si (2b) | 0 | 0 | 0.500 |
| | | | | | O1 (8i) | 0.3074 | -0.3074 | 0.3395 |
| | | | | | O2 (8i) | -0.7910 | 0.7910 | 0.58607 |

[a] (Rignanese & Gonze, 2004)

Table 6 Average static dielectric constant *k*, band gap $E_g$, intrinsic breakdown field $E_{BI}$ and $F_{ED}$ of the $HfO_2$-$SiO_2$ polymorphs

| Compound | Space group (Coord. no. $HfO_2$ & $SiO_2$) | Dielectric constant *k* | | Bandgap $E_g$ (eV) | | $E_{BI}$ (V/nm) | $F_{ED}$ (J/cm$^3$) |
|---|---|---|---|---|---|---|---|
| | | Present | Reported | Present | Reported | | |
| $Hf_1Si_2O_6$ | *P-31m* (6, 6) | 17.53 | | 4.41 | | 1.50 | 174.49 |
| $Hf_2Si_4O_{12}$ | *Cm* (7, 6) | 11.13 | | 5.21 | | 1.77 | 154.84 |
| $Hf_2Si_2O_8$ | *P2/c* (6, 6) | 18.67 | | 4.73 | | 1.61 | 213.71 |
| $Hf_4Si_4O_{16}$ | *I4₁/amd* (8, 4) | 11.43 | 10.64[a] | 5.44 | | 2.58 | 235.42 |
| $Hf_4Si_2O_{12}$ | *C2* (7, 4) | 12.59 | | 4.49 | | 1.53 | 130.11 |
| $Hf_3SiO_8$ | *P-43m* (8, 4) | 18.01 | | 4.42 | | 1.50 | 179.81 |
| $Hf_6Si_2O_{16}$ | *I-42m* (8, 4) | 21.43 | | 4.64 | | 1.58 | 236.56 |

[a] (Rignanese & Gonze, 2004)

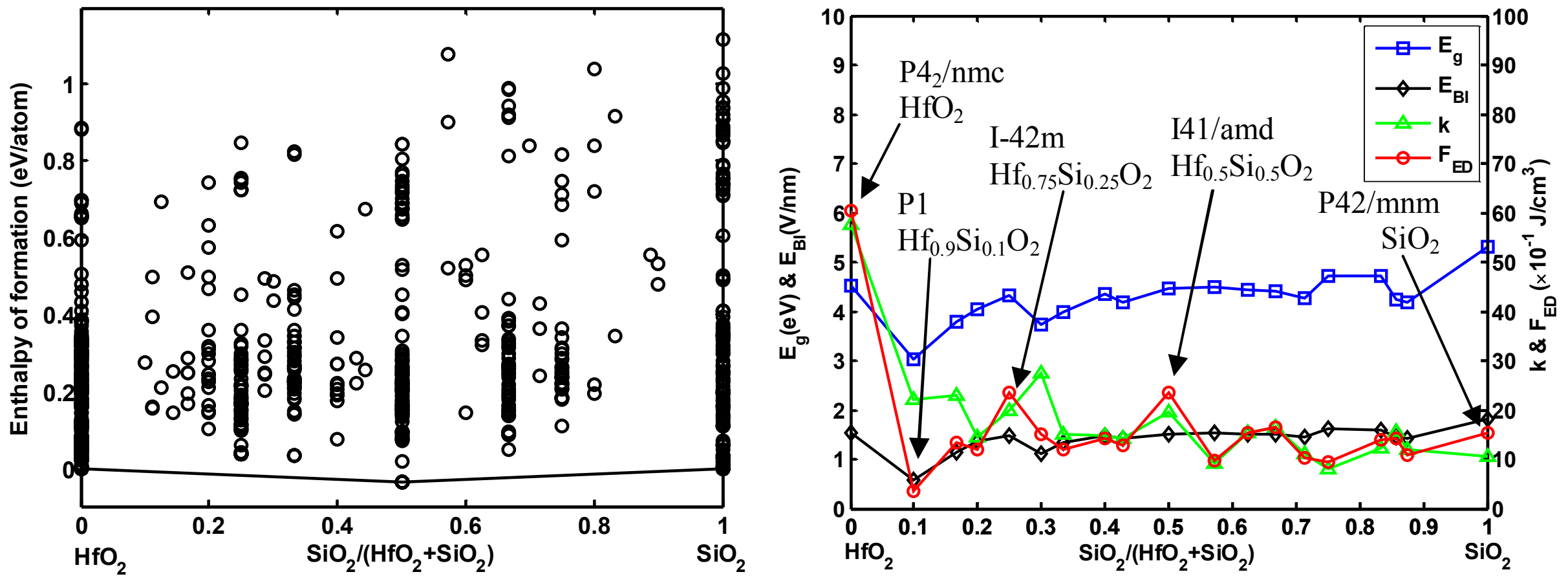

(a) (b)

Fig. 4 The composition of $HfO_2$-$SiO_2$ *vs* (a) enthalpy of formation (from oxides) and (b) physical properties ($k$, $E_g$, $E_{BI}$ and $F_{ED}$), respectively.

## 5. Conclusions

We have proposed a fitness model that can be used as a figure of merit for gate oxide materials and for dielectrics in capacitor applications. This expression contains the static dielectric constant, bandgap, and intrinsic breakdown field. Its use in conjunction with the first-principles global evolutionary algorithm will lead to the discovery of novel dielectrics. This approach has been implemented in the USPEX code and applied to study the $HfO_2$-$SiO_2$ system. Among these structures, we found basic characteristics that can be used as predictors (genes) of high-*k* properties and thus establish clear structure-property relations. Generally, higher coordination numbers correspond to larger dielectric constants. $HfO_2$ (*$P4_2/nmc$*) has ~8 times higher fitness than $SiO_2$ (alpha-quartz). Among the pseudobinary $HfO_2$-$SiO_2$ compounds, $Hf_3SiO_6$ (space group *I-42m*) and $HfSiO_4$ (space group *$I4_1/amd$*) have almost the same highest fitness (energy storage performance), which exceeds that of $SiO_2$ alpha-quartz by ~3 times. This hybrid optimization approach opens up a new avenue for discovering novel high-*k* dielectrics with both fixed and variable compositions, and will speed up the innovative design of materials genome structure-property database.

## Acknowledgements

This work is supported by the Research Fund of the State Key Laboratory of Solidification Processing (No.65-TP-2011, 83-TZ-2013), Foreign Talents Introduction and Academic Exchange Program (No. B08040), the Natural Science Foundation of China (No. 50802076), DARPA (No. W31P4Q1210008 and W31P4Q1310005), NSF (No. EAR-1114313), the Government of the Russian Federation (No. 14.A12.31.0003). The authors also acknowledge the High Performance Computing Center of NWPU for the allocation of computing time on their machines. USPEX code, with options for global optimization of the thermodynamic potential (energy, enthalpy, free energy), hardness, bandgap, dielectric constant, and other properties, is available at: http://uspex.stonybrook.edu.

## References

Aspnes, D. E. & Studna, A. A. (1983). *Physical Review B* **27**, 985-1009.

Batude, P., Vinet, M., Pouydebasque, A., Le Royer, C., Previtali, B., Tabone, C., Hartmann, J. M., Sanchez, L., Baud, L., Carron, V., Toffoli, A., Allain, F., Mazzocchi, V., Lafond, D., Thomas, O., Cueto, O., Bouzaida, N., Fleury, D., Amara, A., Deleonibus, S. & Faynot, O. (2009). *Electron Devices Meeting (IEDM), 2009 IEEE International*, pp. 1-4.

Bersch, E., Rangan, S., Bartynski, R. A., Garfunkel, E. & Vescovo, E. (2008). *Physical Review B* **78**, 085114.

Blöchl, P. E. (1994). *Physical Review B* **50**, 17953-17979.

Campbell, S. A., Gilmer, D. C., Wang, X.-C., Hsieh, M.-T., Kim, H.-S., Gladfelter, W. L. & Yan, J. (1997). *Electron*

*Devices, IEEE Transactions on* **44**, 104-109.
Caravaca, M. A. & Casali, R. A. (2005). *Journal of Physics: Condensed Matter* **17**, 5795-5811.
Choi, J. H., Mao, Y. & Chang, J. P. (2011). *Materials Science and Engineering: R: Reports* **72**, 97-136.
Da Silva, J. L. F., Ganduglia-Pirovano, M. V., Sauer, J., Bayer, V. & Kresse, G. (2007). *Physical Review B* **75**, 045121.
Downs, R. T. & Palmer, D. C. (1994). *American Mineralogist* **79**, 9-14.
Gerritsen, E., Emonet, N., Caillat, C., Jourdan, N., Piazza, M., Fraboulet, D., Boeck, B., Berthelot, A., Smith, S. & Mazoyer, P. (2005). *Solid-State Electronics* **49**, 1767-1775.
Glass, C. W., Oganov, A. R. & Hansen, N. (2006). *Computer Physics Communications* **175**, 713-720.
Hybertsen, M. S. & Louie, S. G. (1987). *Physical Review B* **35**, 5585-5601.
Ieong, M., Doris, B., Kedzierski, J., Rim, K. & Yang, M. (2004). *Science* **306**, 2057-2060.
Iwai, H. & Ohmi, S. i. (2002). *Microelectronics Reliability* **42**, 465-491.
Jiang, H., Gomez-Abal, R. I., Rinke, P. & Scheffler, M. (2010). *Physical Review B* **81**, 085119.
Kadoshima, M., Hiratani, M., Shimamoto, Y., Torii, K., Miki, H., Kimura, S. & Nabatame, T. (2003). *Thin Solid Films* **424**, 224-228.
Kawamoto, A., Cho, K. & Dutton, R. (2001). *Journal of Computer-Aided Materials Design* **8**, 39-57.
Kingon, A. I., Maria, J.-P. & Streiffer, S. K. (2000). *Nature* **406**, 1032-1038.
Kresse, G. & Furthmüller, J. (1996). *Computational Materials Science* **6**, 15-50.
Kukli, K., Ihanus, J., Ritala, M. & Leskela, M. (1996). *Applied Physical Letters* **68**, 3737-3739.
Lee, B. H., Kang, L., Qi, W.-J., Nieh, R., Jeon, Y., Onishi, K. & Lee, J. C. (1999). *Electron Devices Meeting, 1999. IEDM '99. Technical Digest. International*, pp. 133-136.
Lee, C.-K., Cho, E., Lee, H.-S., Hwang, C. S. & Han, S. (2008). *Physical Review B* **78**, 012102.
Lee, S. J., Luan, H. F., Bai, W. P., Lee, C. H., Jeon, T. S., Senzaki, Y., Roberts, D. & Kwong, D. L. (2000). *Electron Devices Meeting, 2000. IEDM '00. Technical Digest. International*, pp. 31-34.
Lyakhov, A. O. & Oganov, A. R. (2011). *Physical Review B* **84**, 092103.
Lyakhov, A. O., Oganov, A. R., Stokes, H. & Zhu, Q. (2013). *Computer Physics Communications* **184**, 1172-1182.
McPherson, J. W., Kim, J., Shanware, A., Mogul, H. & Rodriguez, J. (2003). *IEEE Transactions on Electron Devices* **50**, 1771-1778.
Momma, K. & Izumi, F. (2011). *Journal of Applied Crystallography* **44**, 1272-1276.
Nahar, R. K., Singh, V. & Sharma, A. (2007). *J Mater Sci: Mater Electron* **18**, 615-619.
Oganov, A. R. & Glass, C. W. (2006). *The Journal of Chemical Physics* **124**, 244704.
Oganov, A. R. & Glass, C. W. (2008). *Journal of Physics: Condensed Matter* **20**, 064210.
Oganov, A. R., Lyakhov, A. O. & Valle, M. (2011). *Accounts of Chemical Research* **44**, 227-237.
Perdew, J. P., Burke, K. & Ernzerhof, M. (1996). *Physical Review Letters* **77**, 3865-3868.
Proffen, T., Page, K. L., McLain, S. E., Clausen, B., Darling, T. W., TenCate, J. A., Lee, S.-Y. & Ustundag, E. (2005). *Zeitschrift für Kristallographie - Crystalline Materials* **220**, 1002-1008.
Quintard, P. E., Barbéris, P., Mirgorodsky, A. P. & Merle-Méjean, T. (2002). *Journal of the American Ceramic Society* **85**, 1745-1749.
Rignanese, G.-M., Detraux, F., Gonze, X., Bongiorno, A. & Pasquarello, A. (2002). *Physics Review Letters* **89**, 117601.
Rignanese, G.-M. & Gonze, X. (2004). *Physical Review B* **69**, 184301.
Robertson, J. (2004). *The European Physical Journal Applied Physics* **28**, 265-291.
Robertson, J. (2006). *Reports on Progress in Physics* **69**, 327-396.
Sasaki, H., Ono, M., Yoshitomi, T., Ohguro, T., Nakamura, S., Saito, M. & Iwai, H. (1996). *Electron Devices, IEEE Transactions on* **43**, 1233-1242.
Shishkin, M. & Kresse, G. (2006). *Physical Review B* **74**, 035101.
Shishkin, M. & Kresse, G. (2007). *Physiacl Review B* **75**, 235102.
Wang, L.-M. (2006). *Microelectronics, 2006 25th International Conference on*, pp. 576-579. Belgrade.
Wilk, G. D., Wallace, R. M. & Anthony, J. M. (2001). *Journal of Applied Physics* **89**, 5243-5275.
Wu, H., Zhao, Y. & White, M. H. (2006). *Solid-State Electronics* **50**, 1164-1169.
Xu, Y.-n. & Ching, W. Y. (1991). *Physical Review B* **44**, 11048-11059.
Yamanaka, T. (2005). *Journal of Synchrotron Radiation* **12**, 566-576.
Zhao, X. & Vanderbilt, D. (2002). *Physical Review B* **65**, 233106.
Zhu, Q., Oganov, A. R., Salvado, M. A., Pertierra, P. & Lyakhov, A. O. (2011). *Physical Review B* **83**, 193410.